\documentclass[traditabstract]{aa}
\usepackage{txfonts}
\usepackage{graphicx}
\usepackage{bm}
\usepackage{natbib}
\usepackage{subfigure}
\usepackage{multirow}

\def\pc{\,\rm{pc}}\def\kpc{\,\rm{kpc}}
\def\K{\,{\rm K}}
\def\Gyr{\,\rm{Gyr}}
\def\dd{{\rm{d}}}
\def\mh{\hbox{[M/H]}}

\def\mas{\,{\rm mas}}
\def\cM{{\cal M}}

\def\ybar{\bar{\mathbf{y}}}
\def\xvec{\mathbf{x}}
\def\s2n{S/N}
\def\Teff{T_{\rm eff}}
\def\nr{\mathcal{R}}

\def\bsigma{\bm{\sigma}}\def\bmu{\bm{\mu}}

\def\singlesize{84mm}

\begin{document}
\title{Distance determination for RAVE stars using stellar models III: The nature of the RAVE
survey and Milky Way chemistry}
\author{B. Burnett\inst{1}
  \and J. Binney\inst{1}
  \and S. Sharma\inst{2}
  \and M. Williams\inst{3}
  \and T. Zwitter\inst{4,5}
  \and O. Bienaym\'e\inst{6}
  \and J. Bland-Hawthorn\inst{2}
  \and K.C. Freeman\inst{7}
  \and J. Fulbright\inst{8}
  \and B. Gibson\inst{9}
  \and G. Gilmore\inst{10}
  \and E.K. Grebel\inst{11}
  \and A. Helmi\inst{12}
  \and U. Munari\inst{13}
  \and J.F. Navarro\inst{14}
  \and Q.A. Parker\inst{15}
  \and G.M. Seabroke\inst{16}
  \and A. Siebert\inst{6}
  \and A. Siviero\inst{3,17}
  \and M. Steinmetz\inst{3}
  \and F.G. Watson\inst{18}
  \and R.F.G. Wyse\inst{8}
}
\institute{Rudolf Peierls Centre for Theoretical Physics, Keble Road, Oxford, OX1 3NP, UK
  \and Sydney Institute for Astronomy, School of Physics A28, University of Sydney, NSW 2006, Australia
  \and Astrophysikalisches Institut Potsdam, An der Sternwarte 16, D-14482, Potsdam, Germany
  \and University of Ljubljana, Faculty of Mathematics and Physics, Jadranska 19, 1000 Ljubljana, Slovenia
  \and Center of excellence SPACE-SI, A\v sker\v ceva cesta 12, 1000 Ljubljana, Slovenia
  \and Observatoire astronomique de Strasbourg, Universit\'e  de Strasbourg, CNRS, UMR 7550, 11 rue de l'universit\'e, 67000, Strasbourg, France
  \and Research School of Astronomy and Astrophysics, Australian National University, Cotter Rd., ACT, Canberra, Australia
  \and Johns Hopkins University, Departement of Physics and Astronomy, 366 Bloomberg center, 3400 N. Charles St.,  Baltimore, MD 21218, USA
  \and Jeremiah Horrocks Institute, University of Central Lancashire,  Preston, PR1 2HE, UK
  \and Institute of Astronomy, University of Cambridge, Madingley Road, Cambridge, CB3 OHA, UK
  \and Astronomisches Rechen-Institut, Zentrum f\"ur Astronomie der Universit\"at Heidelberg, M\"onchhofstr. 12-14, D-69120, Heidelberg, Germany
  \and Kapteyn Astronomical Institut, University of Groningen, Landleven 12, 9747 AD, Groningen, The Netherlands
  \and INAF Osservatorio Astronomico di Padova, 36012 Asiago (VI), Italy
  \and Department of Physics and Astronomy, University of Victoria, P.O. box 3055, Victoria,BC V8W 3P6, Canada
  \and Department of Physics and Astronomy, Faculty of Sciences, Macquarie University, NSW 2109, Sydney, Australia
  \and Mullard Space Science Laboratory, University College London, Holmbury St Mary, Dorking, RH5 6NT, UK
  \and Dipartimento di Astronomia, Universita di Padova, Vicolo dell'Osservatorio 2, I-35122 Padova, Italy
  \and Australian Astronomical Observatory, P.O. box 296, Epping, NSW 1710, Australia
}

\date{\today}

\abstract {We apply the method of Burnett \& Binney (2010) for the
determination of stellar distances and parameters to the internal catalogue
of the Radial Velocity Experiment (Steinmetz et al.\ 2006). Subsamples of stars that
either have Hipparcos parallaxes or belong to well-studied clusters, inspire
confidence in the formal errors. Distances to dwarfs cooler than
$\sim6000\,$K appear to be unbiased, but those to hotter dwarfs tend to be
too small by $\sim10\%$ of the formal errors. Distances to giants tend to be
too large by about the same amount. The median distance error in the whole
sample of 216\,000 stars is $28\%$ and the error distribution is similar for
both giants and dwarfs.  Roughly half the stars in the RAVE survey are
giants. The giant fraction is largest at low latitudes and in directions
towards the Galactic Centre.  Near the plane the metallicity distribution is
remarkably narrow and centred on $\mh=-0.04\,$dex; with increasing $|z|$ it broadens out
and its median moves to $\mh\simeq-0.5$.  Mean age as a function of
distance from the Galactic centre and distance $|z|$ from the Galactic plane
shows the anticipated increase in mean age with $|z|$. }

\keywords{Stars: distances,
  Stars: fundamental parameters,
  Galaxy: abundances,
  Galaxy: disk,
  Galaxy: stellar content,
  Galaxy: structure}
\authorrunning{Burnett et al.}
\titlerunning{Distances to RAVE stars}
\maketitle

\section{Introduction}

In recent years there have been significant advances in our knowledge of the
Galaxy due to a number of large-scale surveys in complementary magnitude
ranges. The Hipparcos mission (\citealt{Hipparcos}) obtained astrometry of
unprecedented precision for a sample of bright stars that was complete only
down to $V\sim8$, and at the other end of the scale we have the Sloan Digital
Sky Survey (SDSS, \citealt{SDSS}), going no brighter than around magnitude
$r=14$ and significantly fainter than $r=18$ (\citealt{Ivezic01}). The sizable gap between
these two studies is filled by the Radial Velocity Experiment (RAVE,
\citealt{RaveDR1}), focusing primarily on magnitudes in the range
$9<I<13$. RAVE is still ongoing, and has provided a number of important
results regarding the structure and kinematics of the Galaxy (see for example
\citealt{Smith07}, \citealt{Seabroke08}, \citealt{Munari09}, \citealt{RAVE_velell}, \citealt{Arnaud10}).

One of the challenges thrown up by recent surveys, concentrating as they do
on relatively distant stars, is that of distance estimation. The astrometric
precision required to measure trigonometric parallaxes to these stars is not
yet available, so we must rely on secondary methods for determining
stellar distances. A useful step in this direction was taken by
\cite{Breddels}, who showed that RAVE data could be used to estimate the
parameters and thus distances for objects in the second RAVE data release.
Their technique was honed by \cite{Zwitter10}, who used spectrophotometric
distances to characterise the RAVE survey.  An alternative approach was
developed by \cite{burnett1} in which our prior knowledge of the Galaxy and
stars is more systematically exploited. In this paper we apply this
technique to $\sim216\,000$ stars
observed by RAVE.

We determine spectrophotometric distances in parallel with other stellar
parameters, in particular metallicity, age and mass, so we are able to study
how the giant/dwarf ratio and the distributions in age and metallicity vary
in the region surveyed by RAVE.  Previous applications of Bayesian inference
to the determination of stellar parameters have focused on single parameters
such as age \citep{Jorgensen}; to our knowledge, this is the first study to
consider all parameters together.

In Section \ref{sec:input} we specify the data on which our distances are
based, and in Section \ref{sec:theory} we summarise the algorithm used to
calculate distances. Section \ref{sec:tests} tests the derived distances by comparing them
with (i) Hipparcos parallaxes and (ii) established cluster distances. The
comparison with Hipparcos parallaxes uncovers slight biases in the distances
to hot dwarfs and giants, respectively. Section \ref{sec:tests} tests the validity of the
formal errors by (i) comparing distances to the same stars derived from
independent spectra, and (ii) comparing our distances with those obtained by
\cite{Zwitter10}. The algorithm determines probability distributions for
several stellar parameters in addition to distance. In Section
\ref{sec:output} we display the distribution of formal errors in metallicity,
age and initial mass, in addition to those in distance. In Section
\ref{sec:SF} we use our distances to explore which regions of the Galaxy are
probed by the RAVE survey, and indicate which regions are predominantly
probed with dwarfs or giants. In Section \ref{sec:param} whether examine how
how mean age and metallicity vary within the probed portion of the Galaxy.
Section \ref{sec:conclusions} sums up.

\section{The input data}\label{sec:input}

Since it first took data in 2003 March, the RAVE survey has taken in excess
of 400\,000 spectra with resolution $R\simeq7500$. A sophisticated reduction
pipeline is required to recover stellar parameters such as $\Teff$, $\log g$
and $\mh$ from this huge dataset. The data pipeline involves several
parameters whose values have to be optimised. Unfortunately, the values that
are optimum for one purpose are not optimum for another. In particular, we
will see in \S\ref{sec:hipparcos} that the settings that are optimised for
dwarfs are sub-optimal for giants, and vice versa, so in this work we use
distinct implementations of the pipeline for stars which have $\log g$
greater than, or less than 3.5, with the split based on the value of $\log g$
returned by the version of the pipeline that is used for the high-gravity
stars.  This is the ``VDR2'' version of the pipeline, which was used for
RAVE's 2nd data release (Zwitter et al. 2008) and also to produce a much
larger data set that was released to the collaboration in 2010 January. The
low-gravity stars are processed with the ``VDR3'' version of the pipelines,
which was used for RAVE's 3rd data release (Siebert et al. 2011) and also to
produce data released to the collaboration in 2010 July.

We consider only data for stars that have spectra with signal-to-noise ratio
$\s2n\ge20$, 2MASS photometry with quoted error in $J-K$ less than $0.3$, and
values from RAVE for $T_{\rm eff}$, $\log g$, $\mh_{\rm raw}$, and [$\alpha$/Fe].
These criteria yield data for $216\,064$ distinct objects.  We have
obtained distances and stellar parameters for these stars.

We neglect the effects of extinction by dust because (a) we use near-infrared
data, (b) most of the sample lies at quite high Galactic latitude, and (c) we
do not have a straightforward method of estimating the reddening of
individual stars.

The metallicities given in the RAVE data releases can be refined using
calibration coefficients determined by comparing the raw RAVE metallicities
with those obtained from high-resolution spectroscopy of a subset of stars.
Each version of the pipeline comes with a calibrations: for VDR2 we have
\citep{RaveDR2}
 \[
\mh = 0.938\mh_{\rm raw} + 0.767[\alpha/\hbox{Fe}] -
0.064\log g + 0.404,
\]
 while for VDR3 we have \citep{RaveDR3}
\[
\mh = 1.094\mh_{\rm raw} + 1.21[\alpha/\hbox{Fe}] -
0.711{\Teff\over5040\K} + 0.763
\]
 In the rest of this paper, all input values of [M/H] are those obtained by
using the appropriate calibration formula above.

The observational errors of each star
depend on the star's signal-to-noise ratio ($\s2n$).  From Fig.~19 of
\cite{RaveDR2} we take the errors on temperature, gravity and metallicity at
$\s2n=40$ to be
 \begin{eqnarray}
  \sigma_{\log T} &=& 0.0434,\\
  \sigma_{\log g} &=& \cases{0.5, &if ${\Teff}<8\,000$\,K;\cr
    0.25 + 0.436 \log \left( \frac{{\Teff}}{8\,000\,{\rm K}} \right), &otherwise;}\\
  \sigma_{\rm [M/H]} &=& 1.07 \log{\Teff} - 3.71. \label{eq:sigy_Z}
\end{eqnarray}
{\bf In the great majority of cases these errors are significantly larger
than those given in Table~4 of \cite{RaveDR3} because the latter are simply
internal errors obtained from repeat observations of the same star. As will
become apparent in Section \ref{sec:hipparcos}, there is a substantial
component of systematic error in the stellar parameters, arising from the way
the observed spectra have been fitted to stellar templates. The results of
Section \ref{sec:hipparcos} show that the errors estimates we use are broadly
correct. 

Four entries in Table 4 of \cite{RaveDR3} show errors in $\log g$ that are
more than 25 percent larger than those we have assumed, all for values of
$\log g=2$ or 2.5. In three of these cases the associated error in $\Teff$ is
slightly larger than the value we have assumed.  If future versions of the
pipeline clearly show significant increases in the errors at intermediate
gravities, the errors used in the distance-finding algorithm should be
modified accordingly.}

We obtain the errors at other values $\s2n\ne40$ from the empirical
scaling given by eqs.~(22) and (23) of \cite{RaveDR2}. Errors on colour and
magnitude were taken from the 2MASS measurements on a star-by-star basis. No
allowance is made for the error in [$\alpha$/Fe]. 

We take $J$ and $K$ magnitudes from the 2MASS catalogue (\citealt{2MASS}),
and we use the Padova isochrones (\citealt{Bertelli}) -- these isochrones are
the only widely available ones that reproduce the red clump effectively
\citep{Zwitter10}.

\section{Theory\label{sec:theory}}

We begin by briefly recapping the formalism developed in \cite{burnett1},
where a method for estimating the value of, and error on, each star's
distance, metallicity, age and mass was presented -- \cite{PontEyer} give a
useful introduction to the general methodology. We take the relevant
observables for each RAVE star to be the logarithm of effective temperature
$\Teff$ and surface gravity $g$, the calibrated metallicity $\mh$,
colour $J-K$ and apparent $J$-magnitude. We combine these into the vector of observables
 \begin{equation}
  \mathbf{y} = \left( \log\Teff, \log g, \mh, J-K, J \right) .
\end{equation}
 Each star is assumed to be characterised by a set of `intrinsic' parameters:
true metallicity $\mh_{\rm t}$, age $\tau$, initial mass $\cM$ and heliocentric distance
$s$, which together form a second vector 
 \begin{equation}
  \xvec = \left( \mh_{\rm t}, \log\tau, \cM, s \right).
\end{equation}
 We assume Gaussian observational errors on each component of $\mathbf{y}$,
thus the measured values $\ybar$ for a star have probability density function
(pdf)
 \begin{equation}
  p(\ybar \,|\, \mathbf{y}(\xvec), \bsigma_y) = G(\ybar,\mathbf{y}(\xvec),
  \bsigma_y),
\end{equation}
 where $\mathbf{y}(\xvec)$ represents the true observables corresponding to
intrinsic stellar values $\xvec$, and for an $n$-tuple $\mathbf{w}$ $G$ is defined to be the multivariate
Gaussian
 \begin{equation}
  G(\mathbf{w},\bmu,\bsigma) \equiv \prod_{i=1}^n \left( 
\sigma_i \sqrt{2\pi}\right)^{-1} \exp \left( - (w_i-\mu_i)^2 / 2 \sigma_i^2 \right).
\end{equation}
 The pdf of a
star's intrinsic parameters $\xvec$ is then conditional upon its observed
values $\ybar$, the observational errors $\bsigma_y$ and the fact $S$ that
the star is observed in the survey in question. We thus seek the posterior
pdf $p(\xvec \,|\, \ybar, \bsigma_y , S )$. It is shown in \cite{burnett1}
that the moments of this distribution are given by 
 \begin{equation}\label{eq:integrate}
  \mathcal{I}_{ik} = \int\dd^4\xvec \, x_i^k \, \phi(\xvec) \, G(\ybar,
\mathbf{y}(\xvec), \bsigma_y ) \, p(\xvec) ,
\end{equation}
 where $\phi(\xvec)$ describes any part of the survey selection function that
cannot be expressed as a function of $\ybar$. This leads to a value for the
expectation of each stellar parameter through
 \begin{equation}
  \left< x_i \right> = \frac{\mathcal{I}_{i1}}{\mathcal{I}_{i0}} ,
\end{equation}
and an uncertainty defined by
\begin{equation}
  \sigma_i = \sqrt{\left(\mathcal{I}_{i2} /\mathcal{I}_{i0}\right) - \left< x_i \right>^2} .
\end{equation}

The data were analysed using the prior of \cite{burnett1}, namely a
three-component Milky Way model of the form 
 \begin{equation}\label{eq:priorofx}
  p(\mathbf{x}) = p(\cM) \sum_{i=1}^3 p_i(\mh) \, p_i(\tau) \, p_i(\mathbf{r}),
\end{equation}
 where $i=1,2,3$ correspond to a thin disc, thick disc and stellar halo
respectively. We assumed the same initial mass function (IMF) for all
three components following \cite{Kroupa} and \cite{Aumer}, namely
 \begin{equation}
  p(\cM) \propto \cases{\cM^{-1.3}&if $\cM<0.5\,$M$_\odot$,\cr
    0.536 \, \cM^{-2.2}&if $0.5\,$M$_\odot \leqslant \cM<1\, $M$_\odot$,\cr
    0.536 \, \cM^{-2.519}&otherwise.}
\end{equation}

 The star-formation rate within the thin disc is assumed to have declined
exponentially with time constant $8.4\Gyr$ \citep{Aumer}. The ages of halo
and thick-disc stars are uncertain. We merely assume that the ages of all
halo stars exceed $10\Gyr$, while those of thick-disc stars exceed $8\Gyr$.
Since the thick disc is generally thought to be younger than most of the
halo, we further assume that no thick-disc star has an age in excess of
$12\Gyr$.  The three components are therefore:

\paragraph*{Thin disc ($i=1$):}
\begin{eqnarray} \label{eq:thindisc}
  p_1(\mh) &=& G(\mh,0, 0.2), \nonumber \\
  p_1(\tau)  &\propto& \exp(0.119 \,\tau/\mbox{Gyr}) \quad \mbox{for $\tau \leqslant 10$\,Gyr,}  \\
  p_1(\mathbf{r}) &\propto& \exp\left(-\frac{R}{R_\dd^{\rm{thin}}} - \frac{|z|}{z_\dd^{\rm{thin}}}  \right);  \nonumber
\end{eqnarray}

\paragraph*{Thick disc ($i=2$):}
\begin{eqnarray}
  p_2(\mh) &=& G(\mh,-0.6, 0.5), \nonumber \\
  p_2(\tau)  &\propto& \mbox{uniform in range $8 \leqslant \tau \leqslant 12$\,Gyr,} \\
  p_2(\mathbf{r}) &\propto& \exp\left(-\frac{R}{R_\dd^{\rm{thick}}} - \frac{|z|}{z_\dd^{\rm{thick}}}  \right); \nonumber
\end{eqnarray}

\paragraph*{Halo ($i=3$):}
\begin{eqnarray}
  p_3(\mh) &=& G(\mh,-1.6, 0.5), \nonumber \\
  p_3(\tau)  &\propto& \mbox{uniform in range $10 \leqslant \tau \leqslant 13.7$\,Gyr,} \\
  p_3(\mathbf{r}) &\propto& r^{-3.39}. \nonumber
\end{eqnarray}
 Here $R$ signifies Galactocentric cylindrical radius, $z$ cylindrical
height and  $r$ is spherical radius. The parameter values are given
in Table~\ref{table:params}.

\begin{table}
  \begin{center}
    \caption{Values of disc parameters used.\label{table:params}}
    \begin{tabular}{cr}
      \hline
      Parameter & Value (pc) \\
      \hline
      $R_\dd^{\rm{thin}}$  & 2\,600 \\[3pt]
      $z_\dd^{\rm{thin}}$  & 300 \\[3pt]
      $R_\dd^{\rm{thick}}$ & 3\,600 \\[3pt]
      $z_\dd^{\rm{thick}}$ & 900 \\
      \hline
    \end{tabular}
  \end{center}
\end{table}

\section{Tests}\label{sec:tests}

In this section we show the results of a number of tests we performed to
check the reliability and consistency of our stellar distances.

\subsection{Hipparcos stars\label{sec:hipparcos}}

 \cite{burnett1} demonstrated the robustness of the technique described
in Section~\ref{sec:theory} on the Geneva-Copenhagen sample (\citealt{GCS2}).
However it is clearly important also to make sure that it functions correctly
on the RAVE data.  Consequently, we now investigate the performance of our
method on the subset of RAVE stars that are in the Hipparcos catalogue, in
its re-reduction by \cite{vanLeeuwen}.

We identify RAVE stars that are in the Hipparcos Catalogue by requiring that
sky positions (after updating the Hipparcos positions to J2000) coincide to
$1.5\,$arcsec, and proper motions coincide to $2\sigma$, where $\sigma$ is
the quadrature-sum of the errors in the proper motions in the RAVE and
Hipparcos catalogues. This process leads to 4582 matches, but these matches
include only 4080 distinct stars; the remaining matches arise from multiple
RAVE observations of the same object.
 
\begin{figure*}
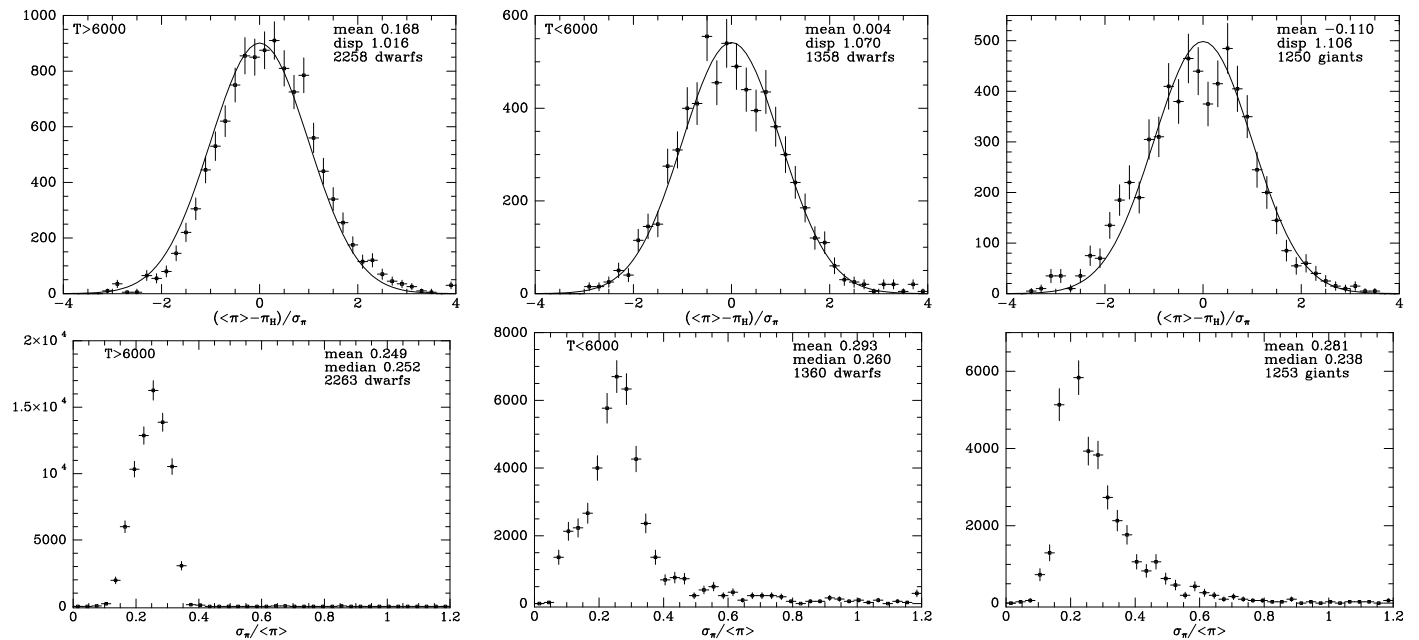

\centerline{\includegraphics[width=.32\hsize]{figures/hot_dwarf_C.ps}\quad
\includegraphics[width=.32\hsize]{figures/cool_dwarf_C.ps}\quad
\includegraphics[width=.32\hsize]{figures/giant_C.ps}}
\centerline{\includegraphics[width=.32\hsize]{figures/hot_dwarf_s.ps}\quad
\includegraphics[width=.32\hsize]{figures/cool_dwarf_s.ps}\quad
\includegraphics[width=.32\hsize]{figures/giant_s.ps}}  
  \caption{Upper panels: distribution of normalized residuals
  (\ref{eq:defsresid}) between our spectrophotometric  parallaxes and
  values from Hipparcos. In this and subsequent histograms the quantity
  plotted vertically is the number of objects in the bin divided by the bin's
  width, and an error bar shows the statistical uncertainty of each point. 
  The mean and dispersion of each distribution are
  given at top right.  A Gaussian of zero mean and unit dispersion is
  over-plotted. Lower panels: the distribution of
fractional uncertainties in the spectrophotometric parallaxes. 
For  the top panel the errors are found by adding
our uncertainty and that of Hipparcos in quadrature; for the bottom
panel the errors are just the spectrophotometric ones.}
  \label{fig:hipresults}  
\end{figure*}

\begin{figure*}
\centerline{\includegraphics[width=.85\hsize]{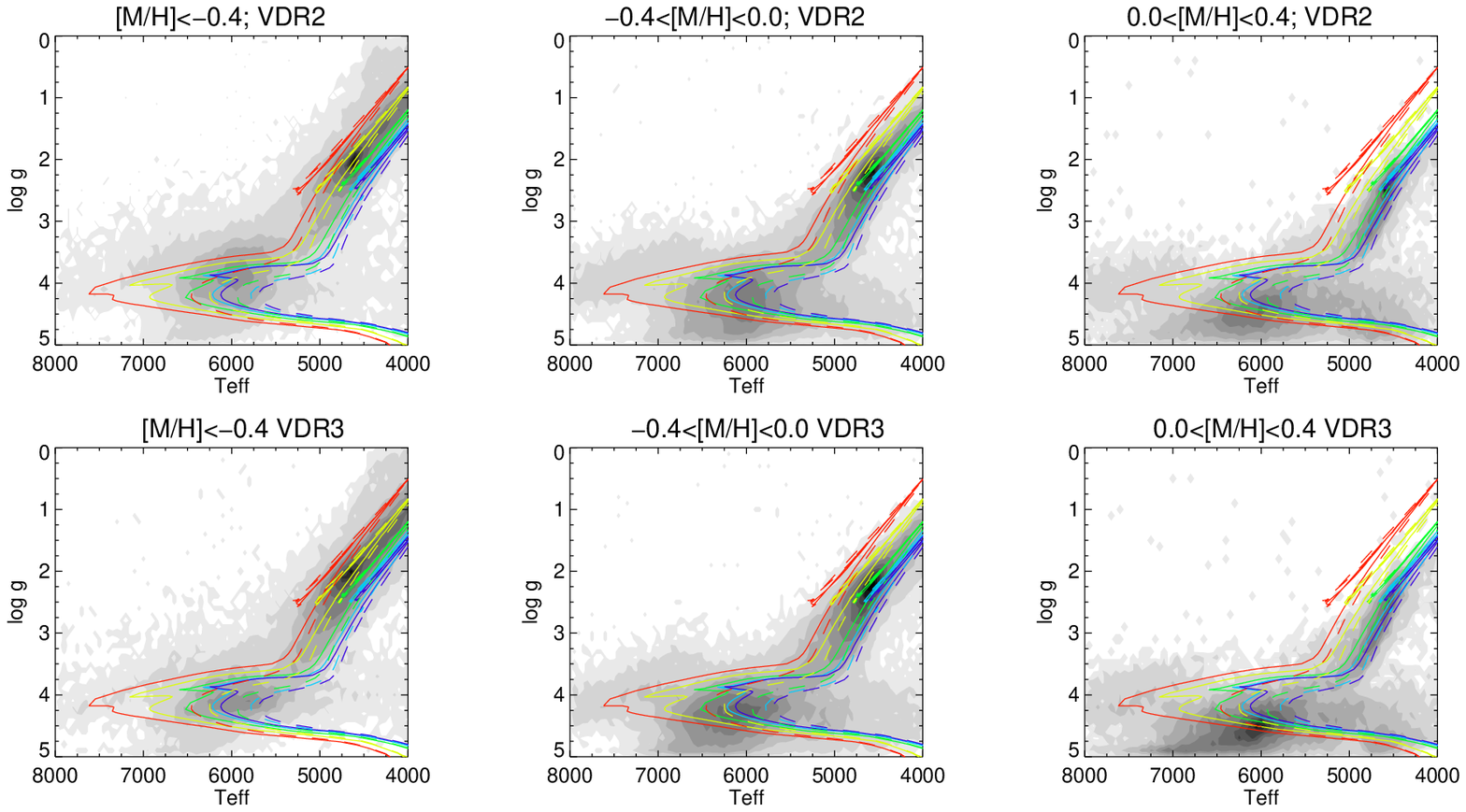}}
\caption{The density of RAVE stars with $\s2n\ge20$ in the $(\Teff,\log g)$
plane together with Padova isochrones for ages $3\Gyr$ (solid) and
$9\Gyr$ (dashed).  The colours indicate calibrated metallicities: $\mh=-1$ red;
$\mh=-0.6$ yellow; $\mh=-0.2$ green; $\mh=0$ light blue; $\mh=0.1$ dark blue.
The top row shows parameters from the VDR2 pipeline, while the lower row
is for the VDR3 pipeline. In each row stars are grouped by metallicity, with
the highest metallicities on the right. Note that in several panels, in the
temperature range $6000\le\Teff\le7000$ the density of stars is high below
the lowest isochrone.  This anomaly is most pronounced in the lower row.}\label{fig:maryfig}
\end{figure*}

Given that the Bayesian method can output parallaxes as easily as distances,
\cite{burnett1} argued for comparisons with the Hipparcos data to be
performed in parallax space. This is what we do in this section. It proves
instructive to compare the stars in three groups: ``giants'' ($\log g<3.5$), ``cool
dwarfs'' ($\log g\ge3.5$ and $\Teff<6000\K$) and ``hot dwarfs'' ($\log g\ge3.5$ and
$\Teff\ge6000\K$).  Fig.~\ref{fig:hipresults} shows the normalised residuals
 \begin{equation}\label{eq:defsresid}
\nr_\varpi\equiv\frac{\langle\varpi\rangle - \varpi_{\rm Hipparcos}}{\sqrt{\sigma_\varpi^2 + \sigma_{\rm Hipparcos}^2}}
\end{equation}
 between the photometric and Hipparcos parallaxes for each of these groups.
In each case the dispersion of the residuals is close to unity, which
confirms the accuracy of the derived errors. For the cool dwarfs the mean
residual is pleasingly close to zero, but the mean residual of the hot dwarfs
is distinctly positive, indicating that the photometric parallaxes tend to be
larger than the trigonometric ones.  Fig.~\ref{fig:maryfig} makes the cause
of this anomaly clear by showing the density of RAVE stars with $ \s2n\ge20$
in the $(\Teff,\log g)$ plane together with isochrones for ages $3\Gyr$
(full) and $9\Gyr$ (broken) and metallicities ranging from $\mh=-1$ (red) to
$\mh=0.1$ (dark blue).  For $\Teff\ga6000\K$ and $\mh>-0.4$ there is a clear
tendency for stars to be concentrated at higher values of $\log g$ than the
isochrones allow. This shortcoming of the data fed to the Bayesian algorithm
leads to the systematic under-estimation of the radii and therefore the
luminosities of stars.  Consequently, the predicted parallaxes are larger
than they should be. Fig.~\ref{fig:maryfig} shows that the displacement of
hot, relatively metal-rich stars to excessive values of $\log g$ is
significantly more pronounced with the VDR3 pipeline (lower row) than with
the VDR2 pipeline, and the stellar parameters from VDR3 yield a mean value of
$\nr_\varpi$ for the hot dwarfs which is as large as $0.273$.

The top right panel of Fig.~\ref{fig:hipresults} shows that the
spectrophotometric parallaxes of giants are systematically too small, so
$\nr_\varpi=-0.11$. When the parameters output by the VDR2 pipeline are used
for giants, we obtain $\nr_\varpi=-0.19$. In fact, the VDR3 pipeline returns
values of $\log g$ that are systematically larger than those returned by the
VDR2 pipeline. For dwarfs the increases are relatively large and for many
stars the VDR3 values are physically implausible. For giants the increases
are smaller and seem to be beneficial.

The mean and dispersion displayed in Fig.~\ref{fig:hipresults} are for the
distributions once we clip outliers, defined as stars with normalized
residuals of modulus greater than four. This excludes eight stars of the
original \mbox{4\,080}. Table~\ref{table:outliers} lists some data for these
objects, one of which has repeat observations. It is notable that stars with
$\log g<4$ are predicted to have smaller parallaxes than Hipparcos measured.
For example, the $\log g$ value of Hipparcos 6075 is strongly indicative of a
giant, so the stellar models predict for it an absolute $J$-magnitude
in the range $M_J \in (-5.7, 0.5)$; combined with its measured $\bar{J}=8.68$
this would imply a parallax in the range $\varpi \in (0.2, 4.2)\mas$.
Consequently the assignment of $\langle\varpi\rangle \in (0.42, 1.04)\mas$ is
eminently reasonable but in strong conflict with the Hipparcos value,
$28.5\mas$. The Hipparcos magnitude of this star is three magnitudes fainter
than its value of $\bar{J}$ from 2MASS.  Either the star is exceptionally
red, or the Hipparcos and 2MASS data relate to different objects.  

Two of the stars with small $\log g$ and under-estimated parallax (Hipparcos
44216 and 46831) lie quite near the Galactic plane and may well be obscured,
which would cause the predicted parallax to be too small. At 
www.rssd.esa.int/SA/HIPPARCOS/docs/vol11\_all.pdf Hipparcos 44216 is listed as
a periodic variable star with amplitude $0.18\,$mag.

\begin{table*}
  \begin{center}
    \caption{Outliers from the analysis of Hipparcos stars. $\nr_\varpi$
    symbolises normalized residuals; parallaxes are measured in mas.\label{table:outliers}}
    \begin{tabular}{r r r c c c c r c c c c c}
      \hline
      & {Hipparcos} & {Analysis} \\
      Hipp ID & ${\varpi}\qquad$ & $\langle\varpi\rangle\qquad$& $l$ & $b$ &
      {${J}$} &{$Hp$} & $T_{\rm eff}$&$\log g$& [M/H] & $\s2n$ \\
      \hline
  6075 & $ 28.51\pm  2.80$ & $  0.73 \pm  0.31$ & 290.9 & -68.4 &   8.68 &  11.67 &   4355 &   2.08 &  -0.32 &   28 \cr
 44216 & $  8.07\pm  1.07$ & $  0.90 \pm  0.15$ & 279.9 & -11.0 &   8.18 &  10.38 &   3967 &   2.42 &  -2.30 &   72 \cr
 46831 & $ 18.10\pm  1.83$ & $  1.57 \pm  3.56$ & 269.1 &   6.0 &   8.97 &  11.37 &   3890 &   3.53 &   0.08 &   21 \cr
 59320 & $ 13.55\pm  1.73$ & $  3.43 \pm  1.26$ & 289.4 &  43.5 &   8.82 &  10.71 &   4724 &   3.57 &   0.14 &   67 \cr
 65142 & $  2.85\pm  1.44$ & $ 38.87 \pm  3.60$ & 315.4 &  54.0 &   7.21 &   9.50 &   6698 &   4.43 &   0.19 &   97 \cr
 73196 & $  3.52\pm  1.03$ & $ 11.12 \pm  1.31$ & 343.7 &  39.0 &   8.09 &   8.93 &   8753 &   4.84 &   0.19 &   92 \cr
 97962 & $ 13.64\pm  1.71$ & $  0.27 \pm  0.05$ &  12.6 & -25.4 &  10.67 &  10.14 &  34684 &   4.95 &  -1.17 &   51 \cr
 97962 & $ 13.64\pm  1.71$ & $  0.54 \pm  0.12$ &  12.6 & -25.4 &  10.67 &  10.14 &  18563 &   4.99 &  -0.59 &   30 \cr
101250 & $  1.17\pm  0.85$ & $ 12.44 \pm  2.18$ &  16.0 & -33.0 &   7.69 &   8.40 &   8130 &   4.26 &   0.63 &   66 \cr
      \hline
    \end{tabular}
  \end{center}
\end{table*}

Hipparcos 65142 has $J-K=0.799$, which is consistent with its being either a
giant or a dwarf, depending on its metallicity. The RAVE catalogue gives
$\mh_{\rm raw}=0.2$ and at this high metallicity a dwarf is consistent with the colour.
The algorithm chooses this solution because $\log g$ is measured to be $4.4$.
On account of the prior, the algorithm returns a probability distribution for
$\mh_{\rm t}$, $0.04\pm0.14$, that is centred on a smaller metallicity.  If
$\mh_{\rm t}$
were near the bottom of this range the star could not be a dwarf and
the predicted parallax would fall to near the Hipparcos value.

The other outliers in Table \ref{table:outliers} with $\log g>4$ all have
$\Teff>7000\,$K, and with one exception have over-estimated parallaxes.  Like
Hipparcos 65142, these stars have probably been assigned values of $\log g$
that are too large and are in consequence predicted to be less luminous than
they really are. The one star that has under-estimated parallax is Hipparcos
97962. \cite{Tsvetkov} have already noted that the spectroscopic distance to
this star exceeds the distance inferred from its parallax by a factor
$\sim37$. They conclude that it is a B9V star and a member of a multiple star
system.  Both of its RAVE spectra imply that it is a very hot star.  The
highest temperature provided by the model isochrones is \mbox{$30\,410\,$K},
so even with allowance for observational error, the temperature from the
higher S/N observation cannot be matched by the program.  Moreover, the
star's metallicity, $-1.17$, falls below that of our lowest-metallicity
isochrone.  Clearly we should exclude from analysis all stars that are so
incompatible with the models.  In practice we excluded all observations with
$\Teff$ above the maximum model temperature (\mbox{$\Teff = 33\,600\,$K}).
Fortunately, this cut removed only two stars from the sample.

\subsection{Cluster stars\label{sec:clusters}}

One other test we can perform involves finding the distances to  RAVE
stars that were identified by \cite{Zwitter10} as lying in clusters with
known distances. Ten of these stars are giants ($\overline{\log g} <
3$), so provide a good test bed for the analysis of such stars, which have
traditionally proved difficult for spectrophotometric distance techniques to
fit (\citealt{Breddels}). The large distances to these stars also
provide a good test of the safety of neglecting reddening. Star
OCL00277\_2236411 has a repeat measurement in the RAVE catalogue and hence we
fitted it twice using both sets of results.

\begin{figure}
  \centerline{\includegraphics[width=\singlesize]{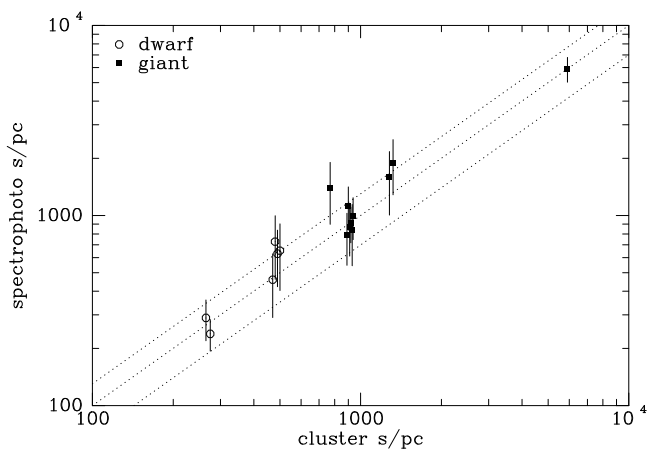}}
  \caption{Estimated distances $\langle s\rangle$ to cluster stars plotted against cluster
distances from the literature.  Dwarf stars are marked by open octagons and
giant stars are marked by filled squares. The dotted lines show equal
distances and distances that differ by 30\%. Stars in the same cluster have
been slightly offset horizontally for clarity.}
  \label{fig:clusters}  
\end{figure}

The results of our fitting of these fifteen stars are shown in
Table~\ref{table:clusters} and Fig.~\ref{fig:clusters}.  All but one of the
stars lie within $1\sigma$ of their literature distances, implying a
reasonable fit to both dwarfs and giants, with no evident bias. The mean
value of the normalised residual $\nr_s=(s-s_{\rm c})/\sigma_s$ is $0.28$,
where $\sigma_s$ is the formal error in our distance $s$. If our distances
were unbiased and the literature distances $s_{\rm c}$ were exact, the
distribution of $\nr_s$ would have zero mean and a dispersion of unity, and
the dispersion will in practice be greater than unity because the literature
distances will contain errors. Hence we expect the mean value of $\nr_s$ to
differ from zero by in excess of $1/\surd{15}$ in addition to any bias in our
distances. Hence the actual mean, $\langle{\nr_s}\rangle=0.28$ is consistent both with
no bias and the level of bias revealed by the Hipparcos sample. The best-fit zero-intercept straight-line fit to the points in
Fig.~\ref{fig:clusters} has a gradient of 1.03, again consistent with no
bias.  

We conclude that in contrast to the finding of \citeauthor{Breddels}, our
results for giants appear to be as reliable as those for dwarfs.  Furthermore
the lack of a demonstrable bias for these more distant stars implies that our
distances are valid despite our neglect of reddening, which is justified a
priori by the use of infrared magnitudes and the arguments of
\cite{Zwitter10}.

\begin{table}
  \begin{center}
    \caption{Expectation values $\langle s\rangle$ of the distances to
    cluster members and their uncertainties both in this paper and in
    \cite{Zwitter10}.\label{table:clusters}} 
    \begin{tabular}{lccc}
      \hline
      & $\hspace{-1cm}$ Cluster &{this paper}&{Z10} \\
      Star ID& $\hspace{-1cm}$ distance (pc) & 
      $\langle s \rangle\pm\sigma_s$ (pc) & $\langle s \rangle\pm\sigma_s$ (pc)\\ 
      \hline
OCL00148\_1373319 &   770 & $1401 \pm 502$ & $1120\pm230$\cr
OCL00147\_1373471 &   490 &  $653 \pm 250$ & $620\pm100$\cr
OCL00277\_2236411 &   490 &  $629 \pm 208$ & $540\pm50$\cr
OCL00277\_2236411 &   490 &  $729 \pm 267$ \cr
OCL00277\_2236511 &   490 &  $459 \pm 169$ & $450\pm60$\cr
  T7751\_00502\_1 &   938 &  $994 \pm 247$ & $1040\pm150$\cr
J000324.3-294849 &   270 &  $238 \pm  45$ & $240\pm20$\cr
J000128.6-301221 &   270 &  $289 \pm  70$ & $240\pm20$\cr
J125905.2-705454 &  5900 & $5908 \pm 881$ & $5500\pm800$\cr
        M67-0105 &   910 &  $788 \pm 241$ \cr
        M67-0135 &   910 & $1126 \pm 290$ & $1130\pm290$\cr
        M67-0223 &   910 &  $867 \pm 256$ & $890\pm140$\cr
        M67-2152 &   910 &  $912 \pm 194$ & $960\pm150$\cr
        M67-6515 &   910 &  $843 \pm 300$ & $990\pm190$\cr
J075242.7-382906 &  1300 & $1896 \pm 611$ & $1290\pm220$\cr
J075214.8-383848 &  1300 & $1588 \pm 583$ & $1740\pm330$\cr
      \hline
    \end{tabular}
  \end{center}
\end{table}

\subsection{Repeat observations}\label{sec:repeats}

A significant number of RAVE stars have been observed more than once.
Although these repeat observations will not reveal systematic errors, they do
provide a valuable test of the quoted errors.

\begin{figure}
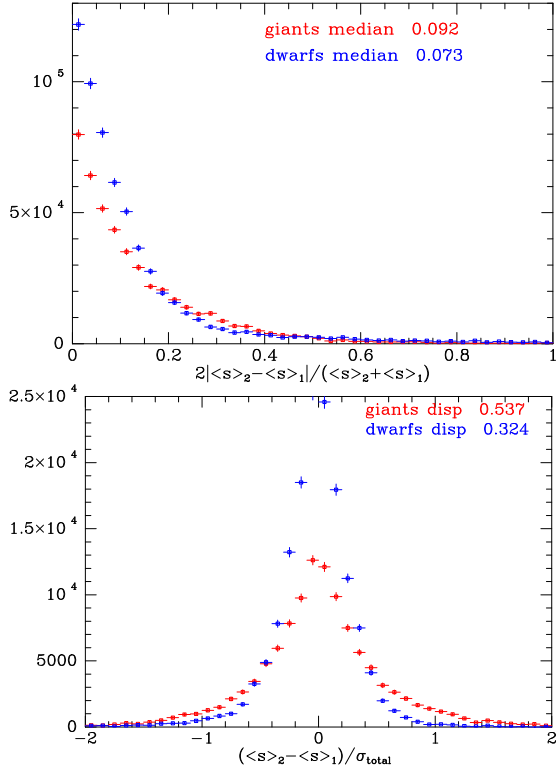

  \centerline{\includegraphics[width=.8\hsize]{figures/repeat1.ps}}
  \centerline{\includegraphics[width=.8\hsize]{figures/repeat2.ps}}
  \caption{Top: the distribution of fractional differences in
distances from repeat observations. Bottom: the distribution of distance
differences divided by the quadrature-sum of the  formal errors of each
distance. As in Fig.~\ref{fig:hipresults}, the quantity plotted vertically is
the number of stars in each bin divided by the bin's width, and the vertical
bars show the statistical uncertainty.}
  \label{fig:repeats}  
\end{figure}

Fig.~\ref{fig:repeats} shows the outcome of this test, using 45\,475 spectra
of $19\,094$ distinct stars.  The top panel shows the distributions of
distance discrepancies divided by the mean distance for giants and dwarfs.
For the giants the median fractional distance residual is 9.2\%, while for
the dwarfs it is only 7.3\%. Taking both populations together, we find that
68.2\% of the points lie at a scatter of below 13.3\%, implying that this may
be a more realistic estimate of the average distance error than the value
implied by the formal errors, 28\% (top left panel Fig.~\ref{fig:rave_errs}
below). The lower panel of Fig.~\ref{fig:repeats} shows the distributions of
normalised residuals, where the normalising factor is the quadrature-sum of
the formal errors on each distance. For both the giants and the dwarfs, the
dispersions of these distributions are considerably smaller than unity, again
implying that the formal errors are excessive, undoubtedly because they are
based on the conservative estimates of the errors on the stellar parameters
derived in \cite{RaveDR2}.

Of course repeat observations will not reveal systematic errors arising from
shortcomings in the isochrones and spectral templates, for example.  However,
the work with Hipparcos and cluster stars described in
Sections~\ref{sec:hipparcos} and \ref{sec:clusters} strongly limits the scale
of systematic errors, which would not be detected by Fig.~\ref{fig:repeats}.
Consequently, we conclude that the mean error in our distances is $\la20\%$,
which is a good level of accuracy for a spectrophotometric technique.

In what follows, we use for stars with several spectra the weighted averages
of parameters from individual spectra, with the weights taken to be the $\s2n$
ratios of the spectra. This leaves us with data for
209\,950 distinct stars.

\begin{figure}
\centerline{\includegraphics[width=.8\hsize]{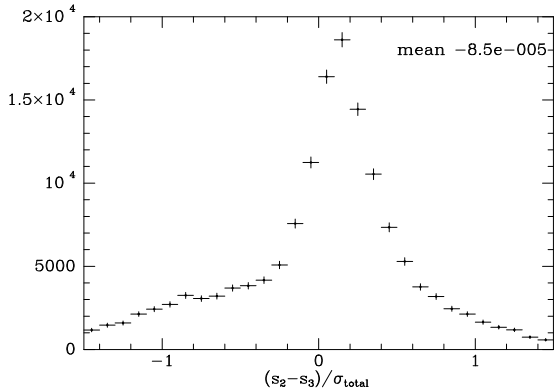}}
\caption{The distribution for $15\,525$ stars with $3.3<\log g<3.7$ of the
differences $s_{\rm VDR2}-s_{\rm VDR3}$ normalised by the quadrature sum of
the formal errors.
}\label{fig:subgs}
\end{figure}

\subsection{Consistency of distances to subgiants}

{\bf Our use of distinct pipelines to assign parameters to stars with $\log g<3.5$
and $\log g>3.5$ raises the question of whether distances are assigned
consistently to stars that have $\log g\simeq3.5$, which may lie on one side
of the dividing line rather than the other merely by virtue of noise.
Fig.~\ref{fig:subgs} addresses this concern by comparing the VDR2 and VDR3
distances for the $15\,525$ stars with $\log g$ in $(3.3,3.7)$. It shows the
distribution of $s_{\rm VDR2}-s_{\rm VDR3}$ normalised by the quadrature sum
of the formal errors in each distance. The mean of the distribution is
extremely small ($-9\times10^{-5}$) although the mode of the distribution lies
near $0.1$. It is to be expected that the higher values of $\log g$ returned
by VDR3 should yield the smaller distances so we expect the peak in
Fig.~\ref{fig:subgs} to lie at a positive value of the ordinate. In view of
the small value of the mean of the distribution, using one pipeline rather
than the other for stars in the transition region cannot be said
systematically to bias the derived distance.}

\subsection{Comparison with \cite{Zwitter10}}

It is interesting to compare our distances with those derived from the same
spectra by \cite{Zwitter10}, which gives distances obtained from the VDR3
pipeline with three different isochrone sets.  Here we consider only the
distances Zwitter et al.\ obtained from Padova isochrones. We use the ``new,
revised'' distances described in the note added to \cite{Zwitter10} in proof;
these distances use the less conservative error estimates to which the
analysis of \cite{RaveDR3} gives rise. Our distances continue to be based on
the older, more conservative error estimates of \cite{RaveDR2}.

The final column of Table~\ref{table:clusters} shows the distances  Zwitter et
al.\ find for the cluster stars. The formal errors on these distances are
smaller than ours on account of the less conservative errors that Zwitter et
al.\ adopted for the input parameters. Otherwise the agreement with our
distances is excellent. In the case of OCL00148\_1373319, for which the
difference between our distance and the literature value is largest
$(1.25\sigma)$, the Zwitter et al.\ distance lies closer to the literature
value than our distance does, although it is still larger than expected by
$1.5\sigma$. This star lies at Galactic latitude $b=3.6^\circ$, while most
RAVE stars lie at much higher latitudes; only 5622 of the spectra under
consideration come from $|b|<6^\circ$. The distances derived for
OCL00148\_1373319 from RAVE data may be too large because the star is
significantly obscured and we have neglected obscuration.

The upper panel of Fig.~\ref{fig:zwitter} shows the distributions of the
normalized residuals between our distances and those of Zwitter at al.\ when
all stars are taken together.  The mean normalised residual is pleasingly
small, $0.02$. The lower panel shows the distribution of normalised residuals
for giants and dwarfs taken separately. The dwarfs have quite a narrow
distribution of residuals (dispersion 0.52) but they are offset from zero by
0.14, implying that the Zwitter et al. distances to dwarfs are systematically
smaller than ours. We saw in \S\ref{sec:hipparcos} a tendency for our
parallaxes for hot dwarfs to be larger than those measured by Hipparcos, so
for these stars our distances are already too small. Thus the Zwitter et al.
distances are offset from ours in the opposite direction from what one would
expect if they were more accurate than ours. This result
undoubtedly reflects the fact that all the Zwitter et al.\ distances are
based on the VDR3 pipeline, which as Figure \ref{fig:maryfig} attests,
over-estimates $\log g$ for hot dwarfs.  

At $0.52$ the dispersion of the normalised residuals of the dwarfs in
Fig.~\ref{fig:zwitter}  is substantially smaller than unity but
larger than the dispersion of repeat observations of dwarfs
(Fig.~\ref{fig:repeats}). This situation is what one would expect given that
the Zwitter et al.\ distances and ours derive from the same data but
processed with
different versions of both the reduction pipeline and the algorithm used to
extract distances from stellar parameters.

In Fig.~\ref{fig:zwitter} the distribution of normalised residuals between the
Zwitter et al.\ distances for giants and ours is quite wide (dispersion
$0.86$) and offset in the opposite direction to the distribution of dwarfs:
the Zwitter et al. distances for giants tend to be larger than ours.  Since
the top right panel of Fig.~\ref{fig:hipresults} implies that our distances
for giants are already larger than the Hipparcos parallaxes imply, the
implication is that the Zwitter et al.\ distances for giants are less
accurate than ours.  

Whereas in Fig.~\ref{fig:zwitter} there is a
contribution to the distribution of normalised residuals for dwarfs from the
different pipelines used (VDR2 versus VDR3), there is no such contribution
to the broader distribution of residuals for giants, and  the entire width of the
giant distribution derives from the different algorithms used to extract
distances from a given set of stellar parameters. 

\begin{figure}
 \centerline{\includegraphics[width=.8\hsize]{figures/Zwitter0.ps}}
 \centerline{\includegraphics[width=.8\hsize]{figures/Zwitter1.ps}}
  \caption{Comparison of our results with those of \cite{Zwitter10}.
Top panel: all $215\,167$ stars with two distances;  lower panel: stars with $\log
g\ge3.5$ or $<3.5$ grouped separately. The statistical uncertainties can be
seen to be 
smaller than some of the points.}
  \label{fig:zwitter}  
\end{figure}

\begin{figure*}
\centerline{\includegraphics[width=.8\hsize]{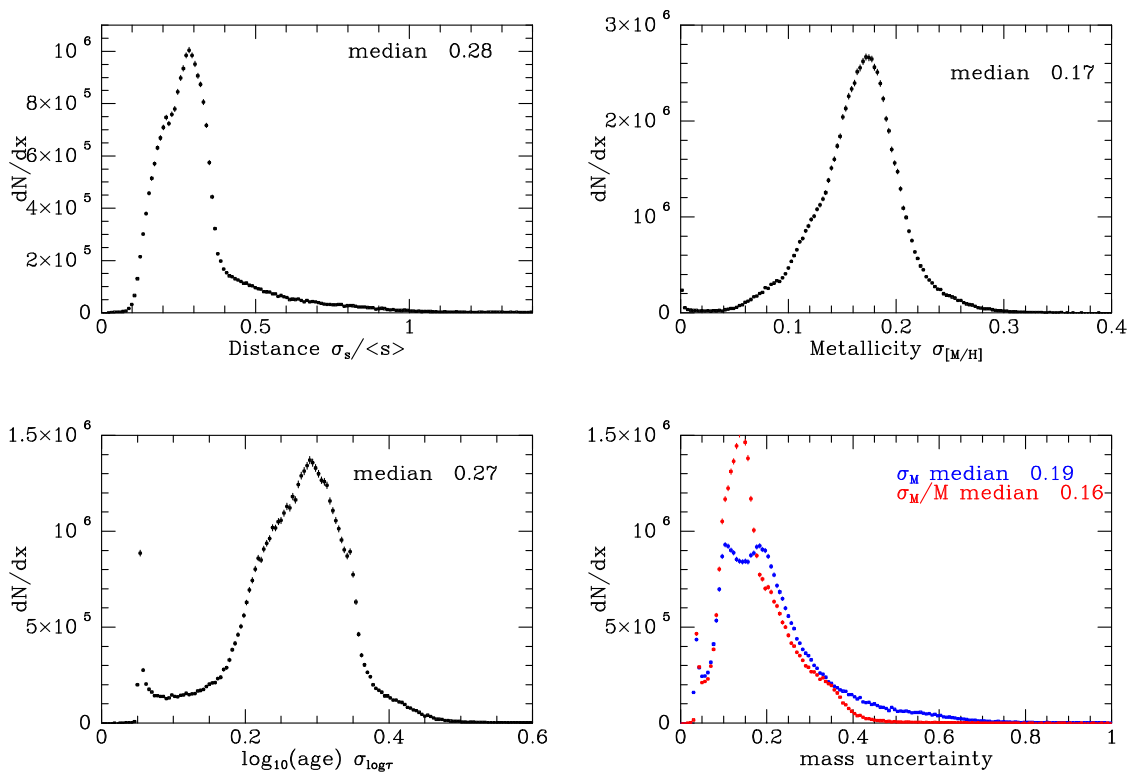}}
  \caption{Distributions of output errors for all four  stellar parameters:
  distance $s$ (top left); metallicity (top right); age (lower left); mass
  (lower right).}
  \label{fig:rave_errs}
\end{figure*}

In fact the residual distribution can be understood in terms of of the
different priors used here and in \citeauthor{Zwitter10}: whereas our prior
recognises the existence of three components in the Galaxy,
\citeauthor{Zwitter10} used a simple prior involving an IMF and a
magnitude-dependent effect to represent the probability of a star entering a
magnitude-limited sample under the assumption of constant volume density.
First, our prior incorporates the spatial inhomogeneity of the Galaxy's
discs, which pulls stars towards smaller distances, particularly at high
Galactic latitudes. The effect on dwarf stars can be expected to be rather
small, but the effect on giants is more marked: in order to fall into the
survey's magnitude limits (which include both low- and high-magnitude cuts),
a giant would have to be at a reasonable distance from the Sun, at which
point (due to RAVE's range of Galactic latitudes) the disc's structure begins
to play a significant role. This accounts for the leftwards wing of the red
histogram  in the
bottom panel of Fig.~\ref{fig:zwitter}.

Second, we also have a prior on stellar age that favours older ages. This
prior increases the likely luminosity of a star of given initial mass over
the most probable luminosity derived from the prior of
\citeauthor{Zwitter10}. Consequently, we identify a significant number of
stars as giants that \citeauthor{Zwitter10} considered to be dwarfs or
subgiants. These stars appear as a noticeable positive wing in the red
histogram of Fig.~\ref{fig:zwitter}.

Aside from these substantial differences in approach, the different IMF and
metallicity distributions taken in our study and that of
\citeauthor{Zwitter10} make the small remaining spread seen in
Fig.~\ref{fig:zwitter} eminently reasonable.

\section{Output precisions}\label{sec:output}

Fig.~\ref{fig:rave_errs} shows the distribution of formal errors in each
stellar parameter for the whole RAVE sample. Comparing these distributions
with those in figs.~3 and 10 of \cite{burnett1}, we see that our precisions
are noticeably higher than forecast. The median formal error in distance is
28\%, and in \S\ref{sec:repeats} we saw that repeat observations suggested
that the random errors are likely only half as large. The median input error
in metallicity from the catalogue is $0.237$, while the top right panel of
Fig.~\ref{fig:rave_errs} shows that the median output error
is $0.17$, so use of the prior diminishes the uncertainty by 29\%. This
reduction is consistent with the small scatter in Fig.~1 of \cite{burnett1}.

\section{The selection function}\label{sec:SF}

Even though RAVE has one of the simplest and best-defined selection criteria
of any large survey of the Galaxy, selection is based on colour and magnitude
on the sky, and it is a non-trivial exercise to compute the resulting
fraction of the stars in a given location in space that will be in the
catalogue. Until these fractions are better determined, we cannot infer
spatial densities of stars in the Galaxy from counts of stars in the RAVE
catalogue. Hence at this stage we are restricted to three lines of enquiry:
(i) what stars does the survey capture?, (ii) how do the distributions of
stellar parameters vary with position in the Galaxy?, and (iii) how do the
Galaxy's kinematics vary with position, age and metallicity?  In this section
we focus on the first two lines of enquiry. Our distances and other parameters
will be used elsewhere to study the Galaxy's kinematics.

\begin{figure}
  \centerline{\includegraphics[width=.8\hsize]{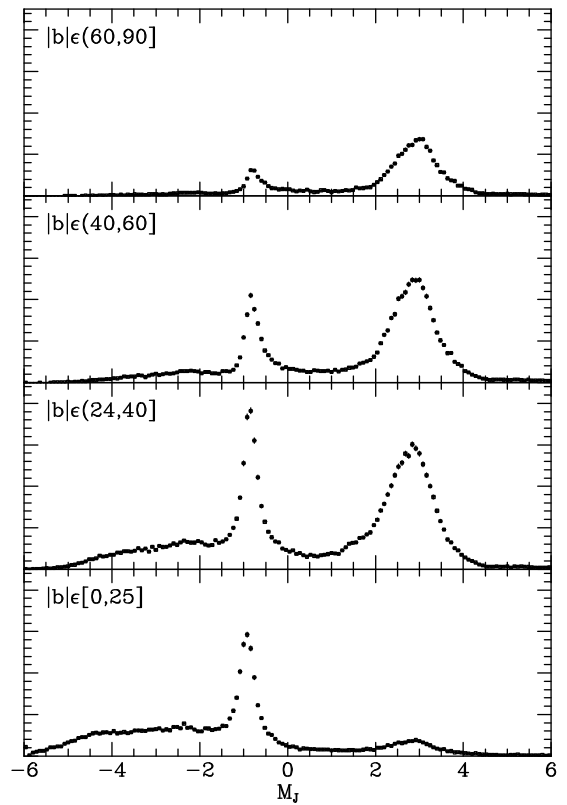}}
  \caption{The distribution of derived absolute magnitudes in slices of
Galactic latitude ($|b|$). The same vertical scale is used for all four
panels so one gets an impression of the latitude-distribution of the entire
sample.}
  \label{fig:MJ_b}
\end{figure}

\begin{figure}
  \centerline{\includegraphics[width=.8\hsize]{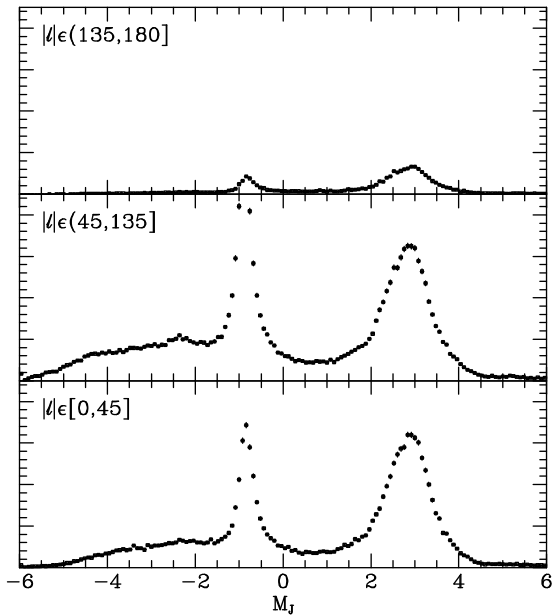}}
  \caption{The distribution of derived absolute magnitudes in ranges of
Galactic longitude ($l$). The same vertical scale is used for all three
panels so one gets an impression of the longitude-distribution of the entire
sample.}
  \label{fig:MJ_l}
\end{figure}

In Fig.~\ref{fig:MJ_b} we show the distribution of absolute $J$-magnitudes in
different slices of Galactic latitude.  The distributions are strongly
bimodal: the red clump produces a narrow peak at $\langle M_J \rangle \sim
-1$ and turnoff stars a broader peak around $\langle M_J \rangle \sim 3$. At
$|b|<25^\circ$ clump stars completely dominate, while the turnoff stars
outnumber giants above $|b|\sim40^\circ$. This progression reflects the
steepening gradient in the density of stars along the line of sight with
increasing $|b|$ since clump stars must have a distance modulus of more than
8, and a distance $\ga400\pc$, to enter the survey, and towards the pole
there are many fewer stars at such distances than nearby dwarfs. In the
absence of a steep gradient along the line of sight, clump stars dominate the
survey because the latter was designed to select disc giants. Moreover the
fraction of giants at $|b|<25^\circ$ is enhanced by the fact that during most
of the survey a colour cut $J-K>0.5$ has been imposed in the region
$230^\circ<l<315^\circ$, $|b|<25^\circ$ precisely in order to favour the
selection of giants -- elsewhere selection has been on magnitude alone. 

Fig.~\ref{fig:MJ_l} shows how the distribution in absolute magnitude varies
with Galactic longitude. As the argument just given leads one to expect,
giants are less prominent towards the anticentre than towards either the
inner Galaxy or the tangent directions. However, the distributions in $l$
are strongly influenced by the survey's non-uniform sky coverage.  Towards
the Galactic centre, the RAVE fields extend much closer to the Galactic
plane, and in such fields giants will be particularly prominent.

\begin{figure}
  \includegraphics[width=\singlesize]{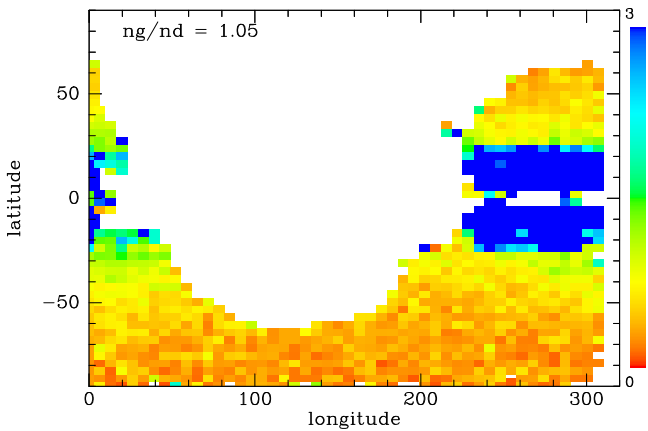}
  \caption{The variation across the sky of the giant/dwarf fraction, with
  giants defined by $\langle M_J \rangle < 1$ and dwarfs by $\langle M_J \rangle
\geqslant 1$. White regions are not sampled by RAVE.}
  \label{fig:giantdwarfratio}
\end{figure}

\begin{figure}
  \includegraphics[width=\singlesize]{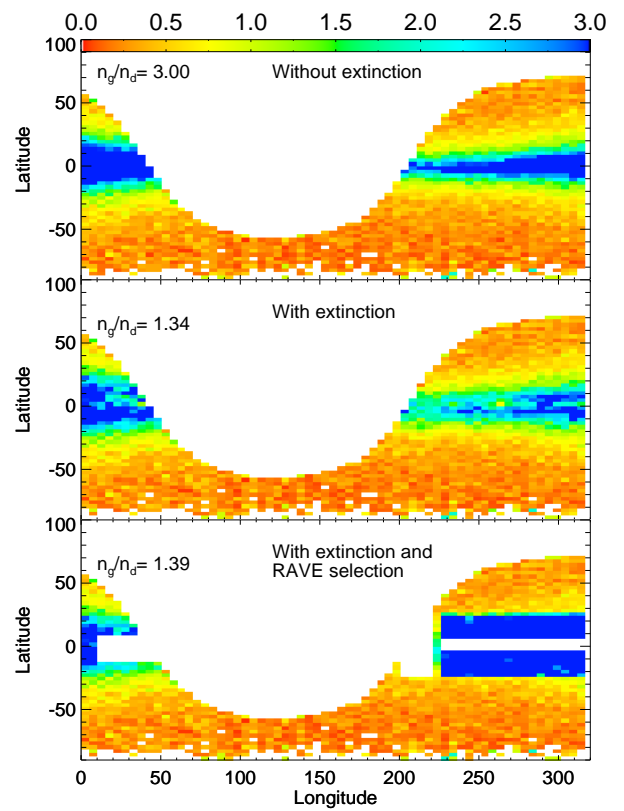}
  \caption{The prediction of the Galaxy modelling code \textit{Galaxia} 
  \citep{Galaxia} for the structure of Fig.~\ref{fig:giantdwarfratio}. The
  top two panels show all stars with $9<I<12$, without (top) and with
  (middle) including Galactic extinction. In the bottom panel low-latitude
  fields have been excluded and for $<225^\circ<l<315^\circ$ only stars with
  $J-K>0.5$ are included.}
  \label{fig:galaxia}
\end{figure}

It is interesting to compare the variation over the sky of the giant/dwarf
ratio with that predicted by the Galaxy modelling code \textit{Galaxia}
\citep{Galaxia} for a survey with RAVE's selection function.
Fig.~\ref{fig:giantdwarfratio} shows the observed giant/dwarf ratio, while
the bottom panel of Fig.~\ref{fig:galaxia} shows the prediction of
\textit{Galaxia} for this figure. The top two panels of
Fig.~\ref{fig:galaxia} show that the predictions of {\it Galaxia} depend
significantly on the model of extinction by dust (which is that of
\citealt{Galaxia_dust}) by showing the giant/dwarf ratio predicted with
(middle panel) and without (top panel) including extinction. The middle and
bottom panels of Fig.~\ref{fig:galaxia} show the impact on the predicted
giant/dwarf ratio of RAVE's handling of low-latitude regions, especially the
imposition of the colour condition $J-K>0.5$ at $225^\circ<l<315^\circ$.  In
both Fig.~\ref{fig:giantdwarfratio} and the bottom panel of
Fig.~\ref{fig:galaxia} the giant/dwarf ratio is above unity only within
$\sim25^\circ$ of the plane, but near the plane it achieves very large values
-- in Fig.~\ref{fig:giantdwarfratio} there are cells with more than 10 dwarfs
and $n_{\rm g}/n_{\rm d}>54$ -- both on account of the length of sight lines
through the disc, and on the imposition of the colour cut $J-K>0.5$, which
enhances the giant fraction.  Consequently, the dark blue region of
Fig.\ref{fig:giantdwarfratio} is heavily saturated.  The numbers in the top
left corner of each panel give the ratio $n_{\rm g}/n_{\rm d}$ of the total
number of giants to dwarfs, which is $1.05$ for the data and $1.39$ for
the model. Given residual uncertainty as to what RAVE's selection function is
at low latitudes, we consider the agreement between the values of $n_{\rm
g}/n_{\rm d}$ from the data and {\it Galaxia} to be satisfactory.

\begin{figure}
  \centerline{\includegraphics[width=\singlesize]{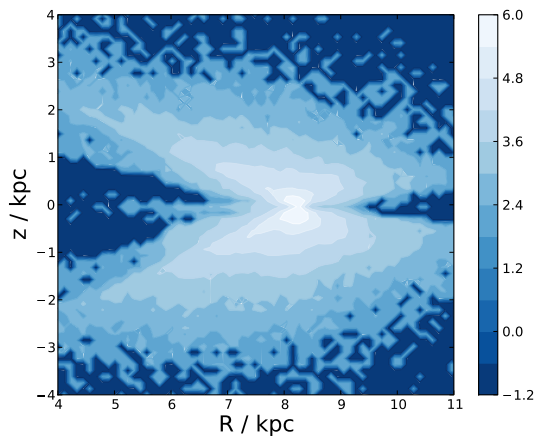}}
  \caption{The distribution of observed stars in the $(R,z)$ plane. Contours
are logarithmically spaced and represent densities in $\log\!\left({\rm
stars\;kpc}^{-2}\right)$.}
  \label{fig:n_Rz}  
\end{figure}

\section{Parameter distributions}\label{sec:param}

What does RAVE tell us about the variation from point to point in the
Galaxy in the distribution of stars over age and metallicity? We must bear in
mind that our prior has a bigger impact on these stellar parameters than on
distances. Consequently, we investigate the effect of our prior on the
results.

\subsection{Region probed by survey}

Fig.~\ref{fig:n_Rz} shows the density of observed stars in the $(R,z)$ plane.
The density seen here is the product of three factors: (i) the intrinsic
density of stars in the Galaxy,  (ii) variation with
distance from the Sun that  follows from the survey's faint and bright
magnitude limits and the stellar luminosity function, and (iii) a bias against
objects in the plane that is driven by a combination of obscuration and the
survey's avoidance of low-latitude fields. Notwithstanding the strong impact
of the biases (ii) and (iii), the basic structure of the Galactic disc is
evident in Fig.~\ref{fig:n_Rz}. 

\begin{figure}
\includegraphics[width=\hsize]{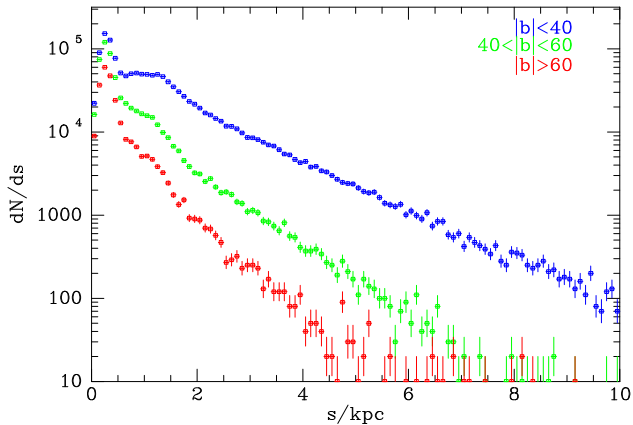}
\caption{Histograms of the distribution in distance of RAVE stars in three
ranges of Galactic latitude.}\label{fig:dhist}
\end{figure}

Regardless of the extent to which the density of observed stars reflects the
Galaxy's intrinsic structure or selection effects, it tells us for which
regions the survey carries useful information. At the solar radius this
extends to $\sim3\kpc$ above and below the plane, and beyond the solar circle
there is a steady but gradual narrowing in the width in $z$ of the surveyed
region with increasing $R$. Towards the centre the surveyed region terminates
at larger values of $|z|$ than towards the anticentre. Fig.~\ref{fig:dhist}
shows the distribution in distance of stars in three ranges of Galactic
latitude: $|b|<40^\circ$, $40^\circ\le|b|<60^\circ$ and $|b|\ge60^\circ$. The
median distances in these three classes are $1.03\kpc$ for $|b|<40^\circ$,
$450\pc$ for $40^\circ\le|b|<60^\circ$ and $372\pc$ for $|b|\ge60^\circ$.

\subsection{Metallicity\label{sec:metallicity}}

\begin{figure}
  \centerline{\includegraphics[width=.9\hsize]{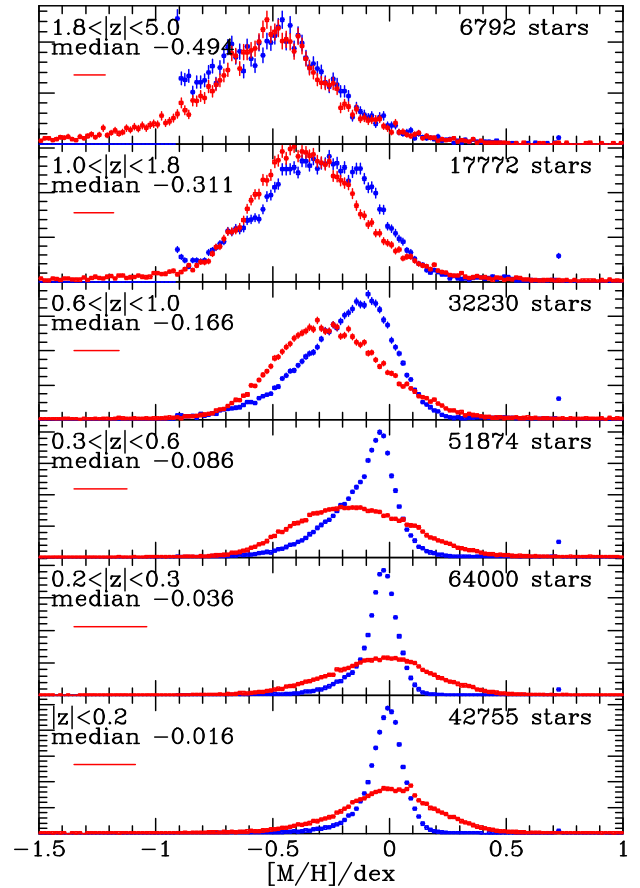}}
  \caption{The  distribution in metallicity at several  distance
  from the Galactic plane. Red points show observed
metallicities, blue points show the output from our analysis. Values of $|z|$  are in kpc, and the median output
metallicity for each slice is displayed. The red
scale bar in each panel represents the median observational error for that
subsample.}
  \label{fig:Z_z}  
\end{figure}

We now look at the variation of mean metallicity with distance from the
Galactic plane. We have two different sets of metallicities to work with,
since each star has both an observed value and that returned by the model
fitting.  We plot both sets of data simultaneously in Fig.~\ref{fig:Z_z}.
Since the model isochrones only cover metallicities down to $\mh=-0.914$,
distributions of output metallicities for $|z|\ge1\kpc$ show a tendency for
stars to pile up near this limit.  At lower heights the output distributions
are distinctly tighter than those observed, and below $\sim300\pc$ they are
displaced to slightly lower $\mh$. The output histograms tend to negligible
values well ahead of the metallicity of the most metal-rich isochrone
($\mh=0.54$), suggesting that there really are
extremely few stars with $\mh_{\rm t}>0.2$. The median value of the observational
error for each slice (estimated via eq.~\ref{eq:sigy_Z} in this paper and
eq.~22 of \citealt{RaveDR2}) is shown as a red scale bar on each panel,
and, given that there are smaller errors on our output metallicities, it can
be seen that, apart from the bias to low metallicity just discussed, the
scale of these errors is very reasonable to explain the difference between
the blue and red distributions in each case.

The progression that would be expected in metallicity is clearly visible as
one moves away from the plane: from a narrow thin-disc distribution at low
$|z|$, there is a gradual shift to a broader thick-disc distribution beyond
around one thin-disc scale height, $\sim0.3\kpc$, moving towards a
significantly lower-metallicity halo distribution as one moves beyond the
thick disc scale height, 0.9\,kpc. Unfortunately, the isochrones we have used
are all too metal-rich for halo stars, so in Fig.~\ref{fig:Z_z} the small
number of halo stars in the RAVE sample cause an un-physical peak in the blue
points at $\mh=-0.9$. The metallicity distribution for
$1\kpc<|z|<1.8\kpc$ is similar to that found by \cite{Bensby07} for
thick-disc stars except that ours extends $\sim0.1\,$dex less far on the
metal-poor side. 

From SDSS data  \cite{Ivezic} concluded that the median metallicity in the
disc decreases from $\mh_{\rm t}=-0.6$ at $|z|=500\pc$ to $-0.8$ beyond
several kiloparsecs. Fig.~\ref{fig:Z_z} implies that at $|z|=500\pc$ the
median disc metallicity is $>-0.1$, and even at $|z|>2\kpc$ it is no smaller
than $-0.5$.

Close to the plane the natural
comparison is with the metallicity distribution within the Geneva-Copenhagen
survey (GCS) of Hipparcos stars \citep{GCS,GCS2}. Unfortunately, the
metallicities of these stars are somewhat controversial. \cite{Fuhrmann08}
finds that high-resolution spectroscopy of a sample of only 185 local F and K
stars implies that the solar-neighbourhood distribution in [Mg/H] covers
$(-0.2,0.2)$, while \cite{Haywood06} argues that the metallicity distribution
of young stars is intrinsically narrow and the spread in measured values of
$\sim0.1\,$dex is dominated by measurement error.  The bottom panel of
Fig.~\ref{fig:Z_z} suggests that near the plane, the intrinsic spread in the
metallicity is indeed narrow.

\begin{figure}
  \centerline{\includegraphics[width=.85\hsize]{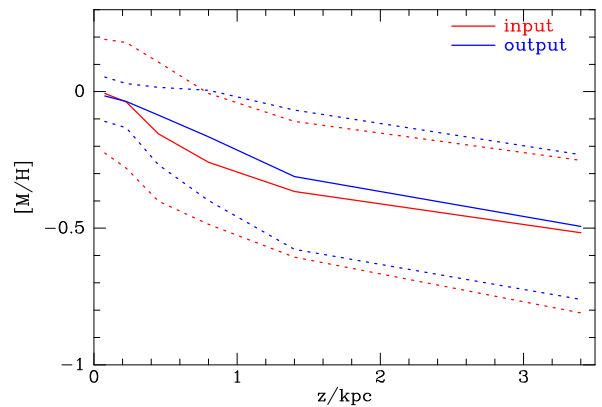}}
  \caption{The variation in  the metallicity distribution
as one moves away from the plane. Red: observed metallicities, blue: output
metallicities. The solid line represents the median of each distribution
and dashed lines show
$1\sigma$ variations from the median.}
  \label{fig:medZ_z}  
\end{figure}

The means of the distributions of input and output metallicities shown in
Fig.~\ref{fig:Z_z} lie close to one another at all values of $|z|$.
Fig.~\ref{fig:medZ_z} makes this fact clear by showing the output
distribution in blue superimposed on the wider input distribution, plotted in
red. The similarities of these means implies that the blue output
metallicities are not merely reproducing our prior.

\begin{figure}
  \centerline{\includegraphics[width=.8\hsize]{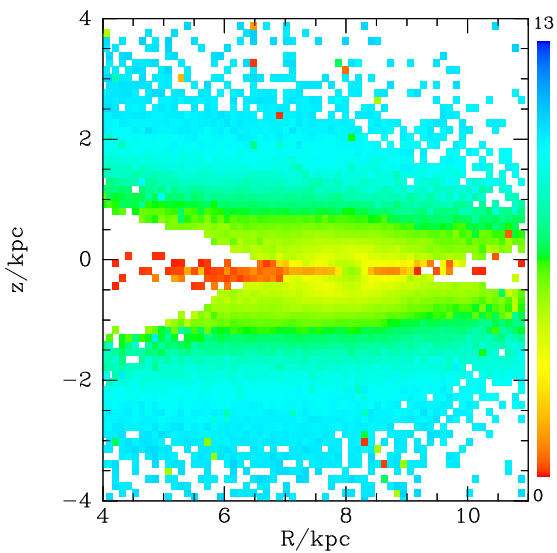}}
  \caption{The distribution of average stellar age in the $(R,z)$ plane.
The colour scale gives ages in  Gyr. Regions for which there is inadequate
data are white.}
  \label{fig:t_Rz}  
\end{figure}

\begin{figure}
  \centerline{\includegraphics[width=.8\hsize]{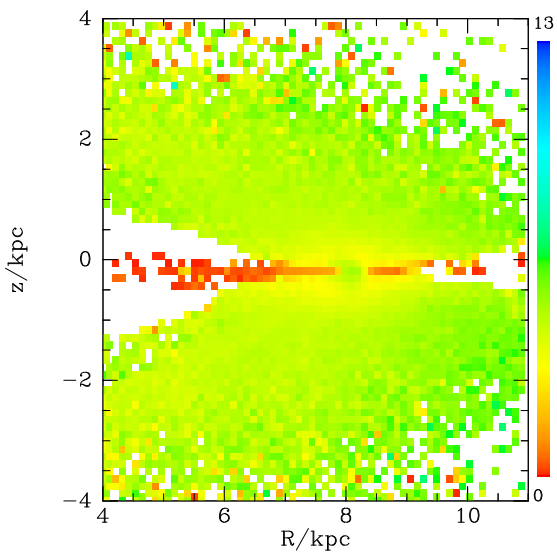}}
  \caption{As Fig.~\ref{fig:t_Rz}, but when the data are analysed with a
  prior that is completely flat in age.}
  \label{fig:t_flatprior}  
\end{figure}

\subsection{Stellar ages}\label{sec:age}

Stellar ages are extremely hard to determine reliably. The most reliable ages
are those obtained from the location in the $(\Teff,L)$ of slightly evolved
main-sequence stars with independently determined distances. Since we do not
know the distances to stars a priori, our ages must be suspect. They are
nonetheless of interest as sanity checks on the performance of the algorithm.
They may even be serve as indicators of possible trends in the data, since
even in the absence of an independent distance estimate for a star, the
spectrum combined with the star's photometry does provide some indication of
its age.

Fig.~\ref{fig:t_Rz} displays a colour-scale plot of the average stellar age
across the $(R,z)$ plane.  To some extent this
distribution reflects our prior, but it is encouraging to see that the map
conforms to our intuition: at small $|z|$ a young disc is dominant. This
young structure dwindles as one moves outwards in $R$ and away from $z=0$.
Far from the plane, the gradient in age becomes small, consistent with these
regions being dominated by an old population that is not strongly
concentrated to the plane. 

Note that shapes of the contours in Fig.~\ref{fig:n_Rz} and
Fig.~\ref{fig:t_Rz} are quite different. This fact is reassuring, for it
tells us that the measured age distribution remains stable even when
the survey picks up only a small fraction of the Galaxy's stars.

Fig.~\ref{fig:t_flatprior} addresses the fear that the age gradient evident
in Fig.~\ref{fig:t_Rz} is an artifact produced by our chosen prior, by
showing the distribution one obtains with a prior that is completely flat in
age from the present back to age \mbox{13.7\,Gyr}; all other elements of the
prior (metallicity, number density, IMF) were unchanged. We see that dropping
the prior causes the mean age for many regions to become $\sim6\Gyr$.  This
happens because with a flat age prior, the age pdfs of stars become broad and
their means shift towards the centre of the permitted range.  Although age is
less strongly correlated with $|z|$ in Fig.~\ref{fig:t_flatprior} than in
Fig.~\ref{fig:t_Rz}, the youngest stars remain concentrated to the plane, and
above the plane there is a tendency for mean age to increase with $R$ at fixed $z$
as we expect if a tapering young disc is superimposed on a broader old
population of thick-disc and halo stars. Thus although the prior is having a
significant effect on the age distribution we recover from RAVE, it is not
entirely responsible for the nice age distribution seen in
Fig.~\ref{fig:t_Rz}.

In Fig.~\ref{fig:t_Rz} the immediate vicinity of the Sun seems to have a
slightly older population than points just above the plane and $\sim1\kpc$
further in or out. The number of stars seen near the plane and $\sim1\kpc$
from the Sun is small (Fig.~\ref{fig:n_Rz}) and the stars we do see have a
high probability of being obscured by dust. The obscuration will select
against low-luminosity stars and thus favour the entry into the catalogue of
hot young stars. It will also affect age determinations, but in an
unpredictable way because $\Teff$ will be changed as well as the broad-band
colours.  Moreover, in low-latitude fields unusual objects were deliberately
targeted by the RAVE survey.  Consequently, the data for low-latitude regions
that lie $\sim1\kpc$ from the Sun are suspect, and the more gradual falloff
in age with height near the Sun is likely to be more representative of the
disc than the steeper gradient seen further away. A countervailing
consideration is that the bright-magnitude limit of RAVE will exclude nearby
young stars and thus bias the nearby data towards older stars. Reassuringly,
when {\it Galaxia\/} is used to simulate Fig.~\ref{fig:t_Rz}, a similar
distribution of ages is produced. In particular, within the plane the mean
age of stars decreases with distance from the Sun for $s\la1\kpc$.


\begin{figure}
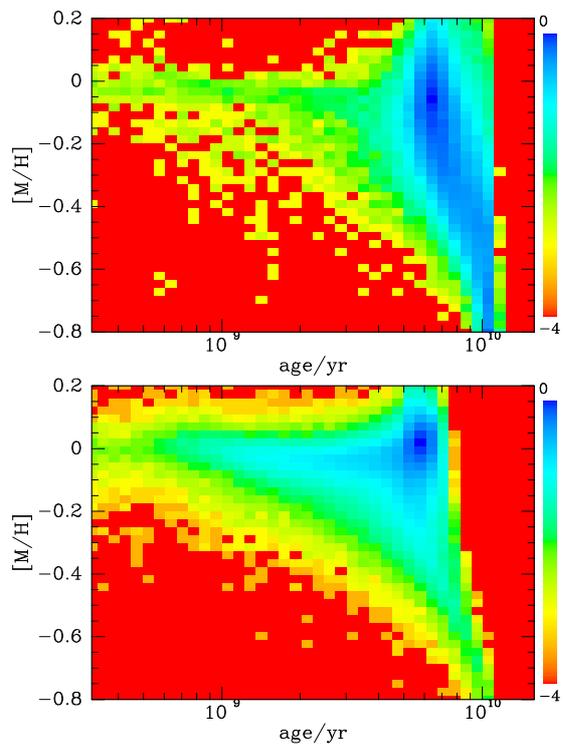

\centerline{\includegraphics[width=.8\hsize]{figures/age_highz.ps}}
\centerline{\includegraphics[width=.8\hsize]{figures/age_lowz.ps}}
\caption{The distribution of stars in the age--metallicity plane at
$|z|>500\pc$ (above) and $|z|<500\pc$ (below). The colour encodes the base-10 logarithm
of stellar density, with red indicating a complete absence of stars. The
values of [M/H] used are the outputs of the distance-finding algorithm.}\label{fig:tauZ}
\end{figure}

\subsection{Age-metallicity relation}

Fig.~\ref{fig:tauZ} shows the distribution of stars in the age--metallicity
plane at $|z|>500\pc$ in the upper panel and $|z|<500\pc$ in the lower panel.
At high $|z|$ we see the expected concentration of thick-disc stars to high
ages with a broad metallicity distribution. The ridge line of the population
clearly shows a rapid increase in metallicity with time, to solar metallicity
at an age of $6\Gyr$.  Nearer the plane, there is a significant population of
young stars and a lower envelope to the distribution that allows for
relatively young stars that have distinctly sub-solar metallicities. The
highest density of thin-disc stars occurs at $\mh_{\rm t}\simeq0$ and age
$\sim6\Gyr$, distinctly younger than the maximum permitted age of disc stars.
And at $\mh_{\rm t}\simeq0$ the density of stars tails off rapidly at ages in excess
of $6\Gyr$. At earlier times only metal-poor stars formed, and the rate at
which they did so appears to have been flat or even increasing with time,
while at later times the star-formation rate must have declined steadily with
time. 

While these trends are interesting and consistent with a plausible scenario
for the chemical evolution of the Galaxy, they should not be assigned much
weight for the reasons given at the start of \S\ref{sec:age}. It is also
worth noting that the plots in Fig.~\ref{fig:tauZ} differ in important
respects from equivalent plots produced by {\it Galaxia}. In particular, the
synthetic plots for $|Z|>500\pc$ do not show a region of high density at
$(\tau=6\Gyr,\mh_{\rm t}>0)$, and, with the expected errors in $\log\tau$,
the distribution of ages at $\mh<-0.4$ is significantly wider than
in Fig.~\ref{fig:tauZ}.

\section{Conclusions}\label{sec:conclusions}

We have derived distances (or parallaxes) to $\sim216\,000$ stars in the RAVE
survey using the Bayesian analysis of \cite{burnett1}. We have checked the
parallaxes and their associated errors against Hipparcos parallaxes, and
conclude that for dwarfs cooler than $\Teff=6000\K$ the parallaxes are
unbiased, but the parallaxes of hotter dwarfs are systematically too large by
$\sim0.1\sigma$ because even the VDR2 pipeline over-estimates the gravities of hot
dwarfs. The parallaxes of giants tend to be too small by $\sim0.1\sigma$.
For all three classes of star, hot dwarfs, cool dwarfs and giants, the
scatter in differences between the spectrophotometric and Hipparcos
parallaxes is consistent with the formal errors on the parallaxes. 

We have checked our distances and our errors against distances to star
clusters, which tend to be beyond the reach of Hipparcos. This rather small
sample of stars, which contains both dwarfs and giants, is consistent with
our distances being unbiased and their errors being accurate. 

RAVE has obtained more than one spectrum for $\sim19\,000$ stars. We have
used this sample to assess our errors by comparing independent distances to
the same object. The scatter in the difference of distances is only half the
quadrature-sum of their formal errors. This is consistent with a significant
contribution to the errors coming from factors, such as defects in the
spectral analysis and the physics of the stellar models, that are the same
for repeated determinations of the distance to a given star.

Comparison of our distances to the spectrophotometric distances of
\cite{Zwitter10} shows that we assign slightly larger distances to dwarfs and
smaller distances to giants than do Zwitter et al. Given the signs of our
deviations from Hipparcos parallaxes, it follows that for both dwarfs and
giants our distances are more accurate than those of Zwitter et al. For
dwarfs this finding can be traced to the use by Zwitter et al.\ of the VDR3
pipeline, which tends to over-estimate $\log g$, especially for hot dwarfs. Our
distances to giants benefit from a more sophisticated prior, which tends to
pull stars to smaller distances. For dwarfs the distribution of normalised
residuals between our distances and those of Zwitter et al.\ is rather
narrow, having a dispersion of only $0.52$, implying that much of the
uncertainty in distance arises from errors in the original data and the
spectral template library, which are common to the two studies. 

Our formal errors are based on the conservative estimates of the errors in
the input data given by \cite{RaveDR2}. The conservative nature of these
errors is confirmed by the analyses of both repeat observations and the distances
of \cite{Zwitter10}. However, the analysis of Hipparcos stars indicates that
by happy chance our formal error budget is just large enough to encompass
external sources of error, such as spectral mis-match and deficiencies in the
stellar models, so our formal errors are close to the final uncertainties in
our distances.

We have examined the distributions of the errors returned by the Bayesian
analysis for distance, metallicity, age and initial mass. The median formal
distance error is $28\%$, the median formal uncertainty in [M/H] is
0.17\,dex, the median formal uncertainty in $\log(\tau)$ is 0.27\,dex and the
median fractional uncertainty in initial mass is $16\%$. These figures show
that by using prior knowledge of the structure of our Galaxy and the nature
of stellar evolution, one can constrain stellar parameters more narrowly than
when each spectrum is considered in isolation.

Data gathered during the pilot part of the RAVE survey has now been released
\citep{RaveDR3}, and the distances we derive from these data are contained in
the release. In view of our results for the Hipparcos stars, it may be useful
to correct the distances of dwarfs with $T>6000\,$K by increasing their
distances by 10\% of their formal errors, and to correct the distances of
giants by decreasing their distances by the same amount. 

Our distances reveal which parts of the Galaxy the RAVE survey probes.
Roughly half the stars in the RAVE catalogue are giants and half dwarfs. The
giant/dwarf ratio varies strongly with Galactic latitude and to a weaker
extent with Galactic longitude. The structure of the variation is accurately
predicted by \textit{Galaxia} when obscuration by dust is included, which
significantly reduces the fraction of giants seen towards the Galactic
Centre.  Although the spatial distribution of RAVE stars reflects the
survey's selection function as well as the intrinsic stellar density of the
Galaxy, it nonetheless reveals the double-exponential structure of the disc.
Moreover, the distribution of stellar ages shows the expected concentration
of young stars towards the plane.

The metallicity distribution evolves systematically with distance from the
plane, being remarkably narrow and slightly sub-solar at $|z|<150\pc$, to much
broader and centred on $\mh\simeq-0.5$ more than $2\kpc$ up. Our results
support the view that observational errors have the biggest impact on the
observed metallicity distribution near the plane.

This work has brought into sharp focus the crucial importance of the pipeline
that extracts stellar parameters from the raw spectra. A valuable upgrade to
the current pipeline would be to force the parameters assigned to a star to
be consistent with models of stellar evolution. This upgrade, which is
mandatory from the perspective of Bayesian inference, would eliminate the
high density of stars in the bottom-right panel of Fig.~\ref{fig:maryfig} at
$\log g\simeq5$ and $\Teff>6000\K$. This study also highlights the need for a
more sophisticated analysis of the errors in the parameters returned by the
pipeline. For example, one would expect the error on $\log g$ to depend on
[M/H] in addition to S/N, which we have had to assume has complete control of
the errors in the input parameters. Moreover, near the turnoff the errors in
$\Teff$ and $\log g$ will be quite strongly correlated, and we have had to
neglect this fact.  All improvements in the pipeline will feed through into
more accurate distances.

The distances and stellar parameters described here provide a basis for
extensive work on the structure, kinematics and dynamics of our Galaxy. Work
on the Galaxy's kinematics is already underway, and papers in this area
will appear shortly. Work on the Galaxy's dynamics requires characterisation
of the intrinsic density distribution of the population(s) whose kinematics
have been measured. Determining those densities involves determination of the
survey's selection function. This is the next major task that must be
accomplished before the RAVE survey can attain its ultimate goals.

\section*{Acknowledgments}

Funding for RAVE has been provided by: the Anglo–Australian Observatory; the
Astrophysical Institute Potsdam; the Australian National University; the
Australian Research Council; the French National Research Agency; the German
Research foundation; the Istituto Nazionale di Astrofisica at Padova; The
Johns Hopkins University; the W.M. Keck foundation; the Macquarie University;
the Netherlands Research School for Astronomy; the Natural Sciences and
Engineering Research Council of Canada; the Slovenian Research Agency; the
Swiss National Science Foundation; the Science \& Technology Facilities
Council of the UK; the US National Science Foundation (grant AST-0908326);
Opticon; Strasbourg Observatory; and the Universities of Groningen,
Heidelberg and Sydney. The RAVE web site is at http://www.rave-survey.org. We
thank members of the Oxford dynamics group for many helpful comments on this
work.  BB acknowledges the support of PPARC/STFC, and JJB the support of
Merton College.
AH acknowledges funding support from the European Research Council
under ERC-StG grant GALACTICA-24027.
\bibliographystyle{aa}
\bibliography{refs}

\end{document}